\documentclass[aps,pra,showpacs,superscriptaddress,preprint]{revtex4}
\usepackage{amsmath}
\usepackage{epsf}
\usepackage{epsfig}

\begin{document}

\title{Time Dependent Entropy of Constant Force Motion}

\author{\"{O}zg\"{u}r \"{O}ZCAN}
\email[E-mail: ]{ozcano@hacettepe.edu.tr}  \affiliation{Hacettepe
\"Universitesi, Fizik E\u{g}itimi B\"ol\"um\"u, Beytepe 06800
Ankara, Turkey}
\author{Ethem AKT\"{U}RK}
\email[E-mail: ]{eakturk@hacettepe.edu.tr}  \affiliation{Hacettepe
\"Universitesi, Fizik M\"uhendisli\u{g}i B\"ol\"um\"u, Beytepe
06800 Ankara, Turkey}
\author{Ramazan SEVER}
\email[E-mail: ]{sever@metu.edu.tr}  \affiliation{Ortado\u{g}u
Teknik \"Universitesi, Fizik  B\"ol\"um\"u, ODTU  Ankara, Turkey}

\begin{abstract}
Time dependent entropy of constant force motion is investigated.
Their joint entropy so called Leipnik's entropy is obtained.  The
main purpose  of this work is to calculate Leipnik's entropy by
using time dependent wave function which is obtained by the
Feynman path integral method. It is found that, in this case, the
Leipnik's entropy increase with time and this result has same
behavior free particle case.

\end{abstract}

\maketitle

\section{Introduction}
The information entropy plays a major role in a stronger
formulation of the uncertainty relations~\cite{ekrem}. This
relation may be mathematically defined by using the
Boltzmann-Shannon information entropy and the von Neumann entropy.
In the literature for both open and closed quantum systems, the
different information-theoretical entropy measures have been
discussed~\cite{Zurek,Omnes,Anastopoulos}. In contrast, the joint
entropy~\cite{Leipnik,Dodonov} can also be used to properties the
loss of information, related to evolving pure quantum
states~\cite{Trigger}. The joint entropy of the physical systems
were conjectured by Dunkel and Trigger~\cite{Dunkel} in which
their systems named MACS (maximal classical states). The Leibnik
entropy of the simple harmonic oscillator was determined not
monotonically increase with time~\cite{Garbaczewski}. In this
work, we give a uniform description of the complete joint entropy
information  of system in motion under a constant force.

This paper is organized as follows. In section II, we explain
fundamental definitions  needed for the calculation. In section
III, we deal with calculation and results for constant force
systems. Moreover, we obtain the analytical solution of Kernel,
wave function in both coordinate and momentum space and its joint
entropy. We also obtain same quantities for constant magnetic
field case. Finally, we present the conclusion in section IV.

\section{Fundamental Definitions}
We consider a classical system with $d=sN$ degrees of freedom,
where N is the particle number and s is number of spatial
dimensions~\cite{Dunkel}. Apart from this, let us describe
$g(x,p,t)=g(x_1,...,x_n,p_1,...,p_d,t)$ which is non-negative,
time dependent phase space density function of system. The density
function is assuming to be normalized to unity,

\begin{equation}
\int dx dp g(x,p,t)=1.
\end{equation}
The Gibbs-Shannon entropy is described by
\begin{equation}
S(t)=-\frac{1}{N!}\int dx dp g(x,p,t)ln(h^{d} g(x,p,t)),
\end{equation}
where $h=2\pi\hbar$ is the Planck constant. Schr\"{o}dinger wave
equation with the Born interpretation~\cite{Born} is given by
\begin{equation}
i\hbar\frac{\partial\psi}{\partial t}=\hat{H}\psi.
\end{equation}
The quantum probability densities are defined in position and
momentum spaces as $|\psi(x,t)|^2$ and $|\tilde{\psi}(p,t)|^2$,
where $|\tilde{\psi}(p,t)|^2$ is given  as
\begin{equation}
    \tilde{\psi}(p,t)=\int\frac{dx
    e^{-ipx/\hbar}}{(2\pi\hbar)^{d/2}}\psi(x,t).
\end{equation}
Leipnik proposed the product function as~\cite{Dunkel}
\begin{equation}
    g_{j}(x,p,t)=|\psi(x,t)|^2|\tilde{\psi}(p,t)|^2\geq0.
\end{equation}
Substituting Eq. (5) into Eq. (2), we get the joint entropy
$S_{j}(t)$ for the pure state $\psi(x,t)$ or equivalently can be
written in the following form ~\cite{Dunkel}
\begin{eqnarray}
    S_{j}(t)&=&-\int dx |\psi(x,t)|^{2}\ln|\psi(x,t)|^{2}-
    \int dp |\tilde{\psi}(p,t)|^2
    \ln |\tilde{\psi}(p,t)|^2-\nonumber\\&-&\ln h^{d}.
\end{eqnarray}
 We find time dependent wave function by means of the Feynman path
integral which has form~\cite{Feynman}
\begin{eqnarray}
K(x'',t'';x',t')&=&\int^{x''=x(t'')}_{x'=x(t')}Dx(t)e^{\frac{i}{\hbar}S[x(t)]}
\nonumber\\&=&\int^{x''}_{x'}Dx(t)e^{\frac{i}{\hbar}\int_{t'}^{t''}L[x,\dot{x},t]dt}.
\end{eqnarray}
The Feynman kernel can be related to the time dependent
Schr\"{o}dinger's wave function
\begin{eqnarray}
K(x'',t'';x',t')=\sum_{n=0}^{\infty}\psi_{n}^{*}(x',t')\psi_{n}(x'',t'').
\end{eqnarray}
The propagator in semiclassical approximation reads
\begin{eqnarray}
K(x'',t'';x',t')=\Big[\frac{i}{2\pi\hbar}\frac{\partial^2}{\partial
x'\partial
x''}S_{cl}(x'',t'';x',t')\Big]^{1/2}e^{\frac{i}{\hbar}S_{cl}(x'',t'';x',t')}.
\end{eqnarray}
The prefactor is often referred to as the Van Vleck-Pauli-Morette
determinant ~\cite{Khandekar,Kleinert}. The $F(x'',t'';x',t')$ is
given by
\begin{eqnarray}
F(x'',t'';x',t')=\Big[\frac{i}{2\pi\hbar}\frac{\partial^2}{\partial
x'\partial x''}S_{cl}(x'',t'';x',t')\Big]^{1/2}.
\end{eqnarray}
\section{CALCULATION AND RESULTS}
\subsection{{\bf Constant Force}}
\begin{figure}[htbp]
\centering \epsfig{file=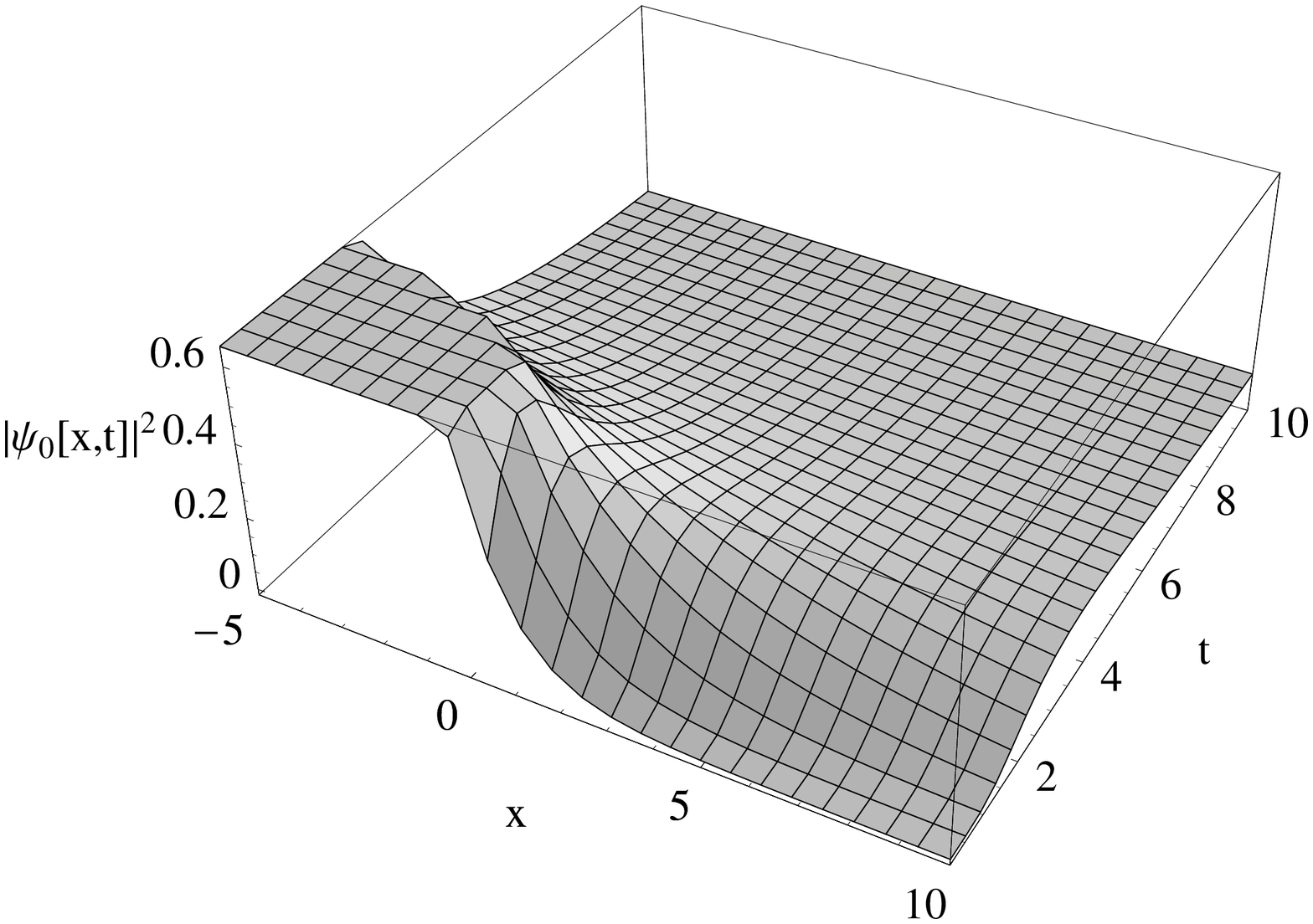, width=10cm,height=10cm}
\caption{$|\Psi(x,t)|^{2}$ versus time and coordinate}\label{eps1}
\end{figure}

The Lagrangian  for present case is
\begin{equation}
L(x,\dot{x},t)=\frac{1}{2}m\dot{x}^2+fx
\end{equation}
The classical path obeys
\begin{equation}
    m\ddot{x}_{cl}=f
\end{equation}
The solution of above equation is
\begin{equation}
    x_{cl}(\tau)=x_{0}+\Big(\frac{x-x_{0}}{t-t_{0}}-\frac{1}{2}\frac{f}{m}(t-t_{0})\Big)\tau+\frac{1}{2}\frac{f}{m}\tau^2
\end{equation}
One obtains for classical action integral along the classical
path~\cite{Feynman}
\begin{eqnarray}
S(x_{cl}(\tau))=\frac{1}{2}m\frac{(x-x_{0})^2}{t-t_{0}}+\frac{1}{2}(x+x_{0})f(t-t_{0})-\frac{1}{24}\frac{f^2}{m}(t-t_0)^3
\end{eqnarray}
and finally, for the kernel
\begin{eqnarray}
K(x'',x';T)&=&\Big[\frac{m}{2\pi i\hbar T}\Big]^{1/2}
\exp\Big[\frac{im}{2\hbar}\frac{(x-x_{0})^2}{T}+\frac{i}{2\hbar}(x+x_{0})fT-\nonumber\\&-&\frac{i}{24\hbar}\frac{f^2}{m}T^3\Big]
\end{eqnarray}
The dependent wave function at time $t>0$
\begin{eqnarray}
 \Psi(x,t)&=&\Big[\frac{1-i\frac{\hbar t}{m\sigma^2}}{1+i\frac{\hbar
 t}{m\sigma^2}}\Big]^{1/4}\Big[\frac{1}{\pi\sigma^{2}(1+\frac{\hbar^{2}
 t^{2}}{m^{2}\sigma^4})}\Big]^{1/4}\exp\Big[-\frac{(x-\frac{p_{0}}{m}t-\frac{ft^2}{2m})^2}{2\sigma^2(1+\frac{\hbar^{2}
 t^{2}}{m^{2}\sigma^4})}\times\nonumber\\&\times&(1-i\frac{\hbar t}{m\sigma^2})\Big]
 \exp\Big[\frac{i}{\hbar}(p_{0}+ft)x-\frac{i}{\hbar}\int^{t}_{0}d\tau\frac{(p_{0}+f\tau)^2}{2m}\Big]
\end{eqnarray}
where $\sigma$ is width of Gaussian curve. The corresponding
probability distribution is
\begin{eqnarray}
|\Psi(x,t)|^2=\Big[\frac{1}{\pi\sigma^{2}(1+\frac{\hbar^{2}
 t^{2}}{m^{2}\sigma^4})}\Big]^{1/2}\exp\Big[-\frac{(x-\frac{p_{0}t}{m}-\frac{ft^2}{2m})^2}{\sigma^2(1+\frac{\hbar^{2}
 t^{2}}{m^{2}\sigma^4})}\Big]
\end{eqnarray}
or
\begin{eqnarray}
|\Psi(x,t)|^2=\Big[\frac{1}{\pi\sigma^{2}(1+\frac{\hbar^{2}
 t^{2}}{m^{2}\sigma^4})}\Big]^{1/2}\exp\Big[-\frac{(x-\frac{p_{0}t}{m}-\frac{ft^2}{2m})^2}{\sigma^2(1+\frac{\hbar^{2}
 t^{2}}{m^{2}\sigma^4})}\Big]
\end{eqnarray}
The probability density in coordinate space is shown
Fig.\ref{eps1}. The probability density in momentum space can be
written easily
\begin{eqnarray}
|\Psi(p,t)|^2=\Big[\frac{\sigma^{2}}{\pi
\hbar^{2}}\Big]^{1/2}\exp\Big[\frac{-\sigma^{2}}{\hbar^{2}}(p+(p_{0}+f
t))^{2}\Big]
\end{eqnarray}

\begin{figure}[htbp]
\centering \epsfig{file=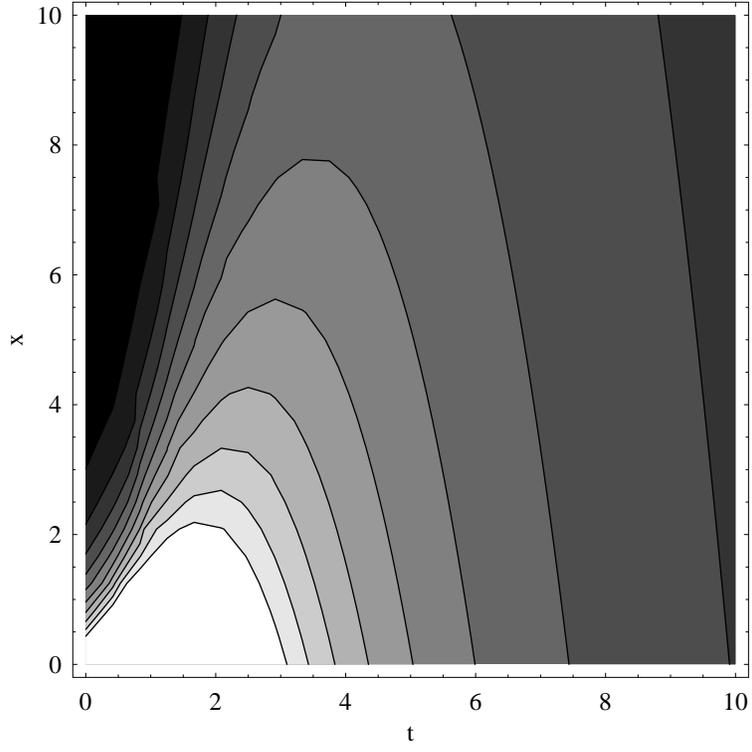,width=10cm,height=10cm}
\caption{The counter graph of $|\Psi(x,t)|^{2}$}\label{eps2}
\end{figure}

\begin{figure}[htbp]
\centering \epsfig{file=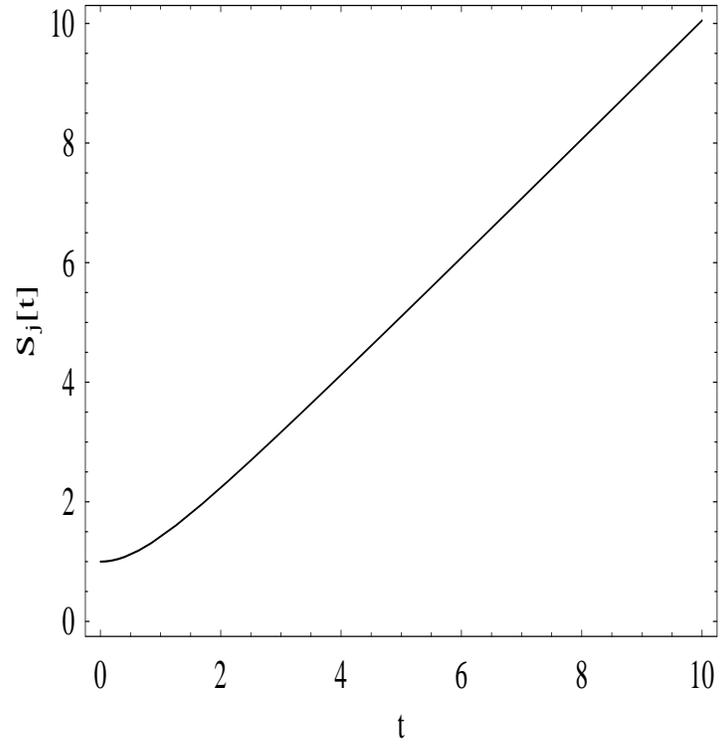, width=10cm,height=10cm}
\caption{The joint entropy of constant forces motion versus
time}\label{eps3}
\end{figure}

The time dependent joint entropy can be obtained from Eq. {6} as
\begin{equation}
    S_{j}(t)=\ln(\frac{e}{2})\sqrt{1+\frac{\hbar^{2}t^{2}}{m^{2}\sigma^{4}}}
\end{equation}
The joint entropy of this system is shown Fig.\ref{eps2} and
Fig.\ref{eps3}. It is important that Eq. {20} is in agreement with
following general inequality for the joint entropy:
\begin{equation}
    S_{j}(t)\geq\ln(\frac{e}{2})
\end{equation}
originally derived by Leipnik for arbitrary one-dimensional
one-particle wave functions.

\section{Conclusion}
We have investigated the joint entropy for constant for motion. We
have obtained the time dependent wave function by means of Feynman
Path integral technique. In this case, we have found that the
joint entropy increase with time and the results harmony prior
studied. The joint entropy has same behavior as free particle
case. This result indicates that the information entropy is
getting increase with time.

\section{Acknowledgements}

This research was partially supported by the Scientific and
Technological Research Council of Turkey.

\newpage
\newpage
\newpage

\end{document}